\documentstyle[12pt]{article}

\begin{document}

\setcounter{page}{1}

\pagestyle{plain} \vspace{1cm}
\begin{center}
\Large{\bf A Lorentz Invariance Violating Cosmology on the DGP Brane}\\
\small \vspace{1cm}
{\bf Kourosh Nozari}\footnote{knozari@umz.ac.ir}\quad\ and \quad {\bf S. Davood Sadatian}\footnote{d.sadatian@umz.ac.ir}\\
\vspace{0.5cm} {\it Department of Physics,
Faculty of Basic Sciences,\\
University of Mazandaran,\\
P. O. Box 47416-95447, Babolsar, IRAN}

\end{center}
\vspace{1.5cm}
\begin{abstract}
We study cosmological implications of a Lorentz invariance violating
DGP-inspired braneworld scenario. A minimally coupled scalar field
and a single, fixed-norm, Lorentz-violating timelike vector field
within an interactive picture provide a wide parameter space which
accounts for late-time acceleration and transition to phantom
phase of the scalar field.
\end{abstract}
\vspace{1.5cm}
\newpage

\section{Introduction}
Theories of extra spatial dimensions, in which the observed universe
is realized as a brane embedded in a higher dimensional bulk, have
attracted a lot of attention in the last few years. In this
viewpoint, ordinary matters and gauge fields are trapped on the
brane but gravitation and possibly non-standard matter can propagate
through the entire spacetime [1,2,3]. In a cosmological perspective,
braneworld scenarios have the capability to explain some of the new
achievements of observational cosmology such as late-time positively
accelerated expansion. Although some of these models predict
deviations from the usual $4$-dimensional gravity at short
distances, the model proposed by Dvali, Gabadadze and Porrati (DGP)
[1] is different in this regard since it predicts deviations from
the standard 4-dimensional gravity over large distances. In fact,
DGP scenario is infra red modification of general relativity.

On the other hand, impacts of Lorentz invariance violation (LIV) on
cosmology have been studied by some authors [4,5]. For instance,
this issue has been studied in the context of scalar-vector-tensor
theories [4]. It has been shown that Lorentz violating vector fields
affect the dynamics of the inflationary models. One of the
interesting feature of this scenario is the fact that exact Lorentz
violating inflationary solutions are related to the absence of the
inflaton potential. In this case, the inflation is completely
associated with the Lorentz violation and depends on the value of
the coupling parameters [5]. One important observation has been made
recently in references [6] which accelerated expansion and crossing
of the phantom divide line with one minimally coupled scalar field
in the presence of a Lorentz invariance violating vector field has
been shown as a result of interactive nature of the model. We know,
an important consequence in quintessence model is the fact that a
single minimally coupled scalar field is not suitable to explain
crossing of the phantom divide line, $\omega=-1$ [7]. However, a
single but non-minimally coupled scalar field is enough to cross the
phantom divide line by its equation of state parameter [8]. It has
been shown that within an interactive picture, a minimally coupled
scalar field in the presence of a Lorentz violating vector field can
evolve to take phantom phase [6]. This happens due to wider
parameter space prepared by inclusion of Lorentz invariance
violating vector field and its interaction with scalar field.

From another viewpoint, currently it is well-known ( based on
various observational data) that our universe has entered the stage
of a positively accelerate expansion around the redshift $z < 1$;
see [9] and references therein. The equation of state (EoS)
parameter $\omega$ responsible for the acceleration of the Universe
has been constrained to be close to $\omega=-1 $. In this regard,
the analysis of the properties of dark energy from recent
observational data mildly favor models of dark energy with $\omega$
crossing $-1$ line in the near past. So, the phantom phase EoS with
$\omega < -1$ is still mildly allowed by observations. Recently,
there have been a number of attempts to realize the phantom phase
EoS, the simplest model which realizes the phantom EoS is provided
by a nonminimally coupled scalar field [7]. Other examples are in
the spirit of braneworld models of dark energy. In these models,
crossing of the phantom divide line and late-time acceleration are
studied extensively ( see for instance [10,11,12]).

With these preliminaries, construction of yet another theoretical
framework which combines braneworld effects and Lorentz invariance
violation to realize positively accelerate expansion and transition
to phantom phase is an interesting challenge. With this motivation,
in which follows, we construct a new dark energy model to realize
crossing of the phantom divide line and explanation of other
observational achievements such as late time acceleration. In this
regard, we study cosmological dynamics of a Lorentz violating
DGP-inspired braneworld scenario. By implementing local Lorentz
violation in a gravitational setting due to the existence of a
tensor field with a non-vanishing expectation value, and then
coupling of this tensor field to gravitational sector and matter (a
scalar field), we study late-time acceleration and transition to
phantom phase of the scalar field. The simplest example of this
approach is to consider a single timelike vector field with fixed
norm. This vector field picks out a preferred frame at each point in
space-time and any matter field coupled to it will experience a
violation of local Lorentz invariance. A special case of this theory
was firstly introduced as a mechanism for Lorentz-violation by
Kostelecky and Samuel in Ref. [13]. In curved spacetime, however,
there is no natural generalization of the notion of a constant
vector field (since $\nabla_\mu u^\nu = 0 $ generically has no
solutions); we must therefore allow the vector field to have
dynamics, and fix its norm by choosing an appropriate action for the
field. We will show that as a result of Lorentz invariance
violation, there is an interacting term in the dynamics of scalar
field which affects cosmological dynamics of the model considerably
and is responsible for transition to phantom phase of the scalar
field.

Nevertheless, one point should be stressed here: we know that DGP
braneworld scenario explains accelerated expansion of the universe
via leakage of gravity to extra dimension [14]. But in this scenario
the EoS parameter of dark energy never crosses the $\omega(z)=-1$
line and universe eventually turns out to be de Sitter phase. On the
other hand, in this setup by incorporating a single scalar field
(ordinary or phantom) on the brane, one can show that EoS parameter
of dark energy crosses the phantom divide line [15]. The question
then arises: why we need to further generalization of this
braneworld setup? One important reason lies in the fact that DGP
setup suffers from ghost instabilities and it is important to add
new ingredients to original setup to overcome this shortcoming. We
thing Lorentz invariance violation may help us to construct a wider
parameter space with potential to overcome this problem.

We begin by an overview of the basic equations of motion for the
most general theory of a fixed-norm vector field $u^\mu$ with a
generalized action $S=S_{Bulk}+S_{Brane}$ where $S_{Brane}\equiv
S_{EH}+S_{\phi}+S_{m}+S_{u}$. Then we generalize this setting to the
model universe with DGP-inspired action and we investigate
cosmological consequences of this setup.

\section{A Lorentz Violating DGP-Inspired Braneworld Scenario}
As it is well known, braneworlds are often studied within the
framework of the $5D$ Einstein field equations projected onto the
$4D$ brane [16]. To study impact of Lorentz invariance violation on
the cosmological dynamics of DGP setup, we consider a vector field
$u^\mu$ along with the extra dimension. So, a local frame at a point
in space-time is inevitably selected as the preferred frame. In
other words, the existence of the brane defines a preferred
direction in the bulk. We study the effects of local Lorentz
violation on the dynamics of the brane by inclusion of this vector
field in the action. This additional field modifies the $4D$
Einstein equations with cosmological implications which we
investigate by studying the resulting Friedmann equation on the
brane.

The action of the Lorentz violating DGP scenario in the presence of
a minimally coupled scaler field and a vector field on the brane can
be written as the sum of two distinct parts
\begin{equation}
S=S_{Bulk}+S_{Brane},
\end{equation}
where $S_{Bulk}$\, and\, $S_{Brane}\equiv
S_{EH}+S_{\phi}+S_{m}+S_{u}$ are defined as follows
\begin{equation}
S_{Bulk}=\int d^{5}x\frac{m^{3}_{4}}{2}\sqrt{-g}{\cal R},
\end{equation}
$$S_{Brane}=\Bigg[\int d^{4}x\sqrt{-q}\bigg(\frac{m_{3}^{2}}{2}
R[q]-\frac{1}{2} q^{\mu\nu} \nabla_{\mu}\phi\nabla_{\nu}\phi
-V(\phi) + m^{3}_{4}\overline{K}+ {\cal{L}}_{m}+$$
\begin{equation}
\Big[-\beta_1 \nabla^\mu u^\nu \nabla_\mu u_\nu-\beta_2 \nabla^\mu
u^\nu \nabla_\nu u_\mu -\beta_3 \left( \nabla_\mu u^\mu
\right)^2-\beta_4 u^\mu u^\nu \nabla_\mu u^\alpha \nabla_\nu
u_\alpha+\lambda \left( u^\mu u_\mu +1
\right)\Big]\bigg)\Bigg]_{y=0},
\end{equation}
where $y$ is coordinate of the fifth dimension and we assume brane
is located at $y=0$.\, $m_4^3$ and $m_3^2$ are fundamental scales in
the bulk and brane respectively. \, $g_{AB}$ is five dimensional
bulk metric with Ricci scalar ${\cal{R}}$, while $q_{\mu\nu}$ is
induced metric on the brane with induced Ricci scalar $R$.\,
$g_{AB}$ and $q_{\mu\nu}$ are related via
$q_{\mu\nu}={\delta_{\mu}}^{A}{\delta_{\nu}}^{B}g_{AB}$.\,
$\overline{K}$ is trace of the mean extrinsic curvature of the brane
defined as
\begin{equation}
\overline{K}_{\mu\nu}=\frac{1}{2}\,\,\lim_{\epsilon\rightarrow
0}\bigg(\Big[K_{\mu\nu}\Big]_{y=-\epsilon}+
\Big[K_{\mu\nu}\Big]_{y=+\epsilon}\bigg),
\end{equation}
and corresponding term in the action is York-Gibbons-Hawking term
[17]. This action is allowed to contain any non-gravitational
degrees of freedom in the framework of Lorentz violating
scalar-vector-tensor theory of gravity. As usual, we assume $u^\mu
u_\mu = -1$ and that the expectation value of vector field $u^\mu$
is $<0| u^\mu u_\mu |0> = -1$ \,[18]. $\beta_i(\phi)$ ($i=1,2,3,4$)
are arbitrary parameters with dimension of mass squared and
$\lambda$ is a Lagrange multiplier. Note that $\sqrt{\beta_{i}}$ are
mass scale of Lorentz symmetry breakdown [4,18,19]. In which
follows, we neglect quartic self-interaction term,  $u^\mu u^\nu
\nabla_\mu u^\alpha \nabla_\nu u_\alpha$. Some cosmological
consequences of this term in the action are studied in Refs. [4,19]
for ordinary $4D$ framework. In this setup, the preferred frame is
selected through the constrained vector field $u^\mu$ and this leads
to the violation of the Lorentz symmetry.\\
The ordinary matter part of the action is shown by Lagrangian
${\cal{L}}_{m}\equiv {\cal{L}}_{m}(q_{\mu\nu},\psi)$ where $\psi$ is
matter field and corresponding energy-momentum tensor is
\begin{equation}
T_{\mu\nu}=-2\frac{\delta{\cal{L}}_{m}}{\delta
q^{\mu\nu}}+q_{\mu\nu}{\cal{L}}_{m}.
\end{equation}
The pure scalar field Lagrangian,\, ${\cal{L}}_{\phi}=-\frac{1}{2}
q^{\mu\nu} \nabla_{\mu}\phi\nabla_{\nu}\phi -V(\phi)$,\,\,  yields
the following energy-momentum tensor
\begin{equation}
 \tau_{\mu\nu}=\nabla_\mu\phi\nabla_\nu\phi-\frac{1}{2}q_{\mu\nu}(\nabla\phi)^2
-q_{\mu\nu}V(\phi).
\end{equation}
The energy-momentum tensor of the Lorentz violating vector field is
defined as usual by
\begin{equation}
T_{\mu\nu}^{(u)}=-2\frac{\delta{\cal{L}}^{(u)}}{\delta
q^{\mu\nu}}+q_{\mu\nu}{\cal{L}}^{(u)}.
\end{equation}
The Bulk-brane Einstein's equations calculated from action (1) are
given by
\begin{equation}
m^{3}_{4}\left({\cal R}_{AB}-\frac{1}{2}g_{AB}{\cal R}\right)+
m^{2}_{3}{\delta_{A}}^{\mu}{\delta_{B}}^{\nu}\left(R_{\mu\nu}-
\frac{1}{2}q_{\mu\nu}R\right)\delta(y)=
{\delta_{A}}^{\mu}{\delta_{B}}^{\nu}\Upsilon_{\mu\nu}\delta(y)
\end{equation}
where $\Upsilon_{\mu\nu}\equiv
T_{\mu\nu}+\tau_{\mu\nu}+T_{\mu\nu}^{(u)}$. From equation (8) we
find
\begin{equation}
G_{AB}={\cal R}_{AB}-\frac{1}{2}g_{AB}{\cal R}=0
\end{equation}
and
\begin{equation}
G_{\mu\nu}=\left(R_{\mu\nu}-
\frac{1}{2}q_{\mu\nu}R\right)=\frac{\Upsilon_{\mu\nu}}{m^{2}_{3}}
\end{equation}
as Einstein's equations in the bulk and brane respectively. The
corresponding junction conditions relating the extrinsic curvature
of the brane to its energy-momentum tensor, have the following form
( see [20] and related references therein)
\begin{equation}
\lim_{\epsilon\rightarrow+0}\Big[K_{\mu\nu}\Big]^{y=+\epsilon}_{y=-\epsilon}
=\frac{1}{m_{4}^{3}}\bigg[\Upsilon_{\mu\nu}-\frac{1}{3}q_{\mu\nu}q^{\alpha\beta}
\Upsilon_{\alpha\beta}\bigg]_{y=0}
-\frac{m^{2}_{3}}{m^{3}_{4}}\bigg[R_{\mu\nu}-
\frac{1}{6}q_{\mu\nu}q^{\alpha\beta}R_{\alpha\beta}\bigg]_{y=0}.
\end{equation}
We start with the following line element to derive cosmological
implications of our model
\begin{equation}
ds^{2}=q_{\mu\nu}dx^{\mu}dx^{\nu}+b^{2}(y,t)dy^{2}=-n^{2}(y,t)dt^{2}+
a^{2}(y,t)\gamma_{ij}dx^{i}dx^{j}+b^{2}(y,t)dy^{2}.
\end{equation}
For such a metric, to solve Einstein equations in the presence of a
fixed-norm vector field, the vector field must respect spatial
isotropy, at least in the background (though perturbations will
generically break the symmetry). Thus the only component that the
vector can possess is the timelike component. Therefore, we take the
constraint $u^{\mu}=(\frac{1}{\cal{N}},0,0,0)$ where $\cal{N}$ is a
lapse function. After performing required algebra, we set $n(0,t)=1$
and ${\cal{N}}=1$ in which follows. The scale of the universe is
determined by $a(y,t)$ and $\gamma_{ij}$ is a maximally symmetric
$3$-dimensional metric defined as
\begin{equation}
\gamma_{ij}=\delta_{ij}+k\frac{x_{i}x_{j}}{1-kr^{2}}
\end{equation}
where $k=-1,0,1$ parameterizes the spatial curvature and
$r^2=x_{i}x^{i}$. We assume that scalar field $\phi$ depends only on
the proper cosmic time of the brane and we adopt the gauge
$b^{2}(y,t)=1$ in Gaussian normal coordinates.\\
Now total energy density and pressure are given as follows
\begin{equation}
\rho_{tot}=\rho_m+\rho_u+\rho_\phi
\end{equation}
and
\begin{equation}
p_{tot}=p_m+p_u+p_\phi.
\end{equation}
Here we assume that energy-momentum tensor of ordinary matter on the
brane has a perfect fluid form with energy density $\rho_m$ and
pressure $p_m$ so that $T_{\mu\nu}=(\rho_m+p_m)N_\mu N_\nu+p_m
q_{\mu\nu}$ where $N_\mu$ is a unit timelike vector field
representing the fluid four-velocity. We also assume a linear
isothermal equation of state for the fluid
$p_m=(\gamma_{m}-1)\rho_m$ that $1\leq\gamma_{m}\leq 2$. Energy
density and pressure of minimally coupled scalar field are given as
follows
\begin{equation}
\rho_{\phi}=\left[\frac{1}{2}\dot{\phi}^{2}+n^{2}V(\phi)\right]_{y=0},
\end{equation}
and
\begin{equation}
p_{\phi}=\left[\frac{1}{2n^{2}}\dot{\phi}^{2}-V(\phi)\right]_{y=0},
\end{equation}
where a dot denotes the derivative with respect to cosmic time\,
$t$. The stress-energy for the vector field also takes the form of a
perfect fluid, with energy density given by [4,5]
\begin{equation}
\rho_u=-3\beta H^2
\end{equation}
and pressure
\begin{equation}
p_u=\beta
H^2\Big[3+2\frac{\dot{H}}{H^2}+2\frac{\dot{\beta}}{H\beta}\Big]
\end{equation}
where we have defined the parameter $\beta$ as
$\beta\equiv(\beta_1+3\beta_2+\beta_3)$ and
$H=\frac{\dot{a}(0,t)}{a(0,t)}$ is the Hubble parameter. In the
absence of vector field, that is, when all $\beta_i = 0$, the above
equations reduce to the conventional ones and in the case $\beta=
const.$, the above equations are lead to the equations given in
[19]. For future references, the total equation of state parameter
defined as $\omega_{tot}=\frac{p_{tot}}{\rho_{tot}}$  or
$$\omega_{tot}=\frac{p_m+p_u+p_\phi}{\rho_m+\rho_u+\rho_\phi}$$
takes the following form
\begin{equation}
\omega_{tot}=\bigg[\frac{p_m+\beta
H^2\Big[3+2\frac{\dot{H}}{H^2}+2\frac{\dot{\beta}}{H\beta}\Big]
+\frac{1}{2n^{2}}\dot{\phi}^{2}-V(\phi)}{\rho_m-3\beta
H^2+\frac{1}{2}\dot{\phi}^{2}+n^{2}V(\phi)}\bigg]_{y=0}.
\end{equation}
Now, the effective Einstein equations on the brane are given as
follows [21]
\begin{equation}
G_{\mu\nu}=\frac{\Pi_{\mu\nu}}{m_{4}^{6}}-{\cal{E}}_{\mu\nu},
\end{equation}
where
\begin{equation}
\Pi_{\mu\nu}=-\frac{1}{4}{\Upsilon}_{\mu\sigma}{{\Upsilon}_{\nu}}^{\sigma}+
\frac{1}{12}{\Upsilon}{\Upsilon}_{\mu\nu}+
\frac{1}{8}g_{\mu\nu}\Big({\Upsilon}_{\rho\sigma}{\Upsilon}^{\rho\sigma}-\frac{1}{3}{\Upsilon}^{2}\Big),
\end{equation}
and
\begin{equation}
{\cal{E}}_{\mu\nu}=C_{IJKL}\,\, \Theta^{I}\,\,\Theta^{K}
{g^{J}}_{\mu}\,\,{g^{L}}_{\nu}
\end{equation}
where $C_{IJKL}$ is five dimensional Weyl tensor and $\Theta_{A}$ is
the spacelike unit vector normal to the brane. Using equation (21)
we find
\begin{equation}
{G^{0}}_{0}=\frac{{\Pi^{0}}_{0}}{m_{4}^{6}}-{{\cal{E}}^{0}}_{0}
\end{equation}
where for FRW universe we have
\begin{equation}
{G^{0}}_{0}=-3\Big(H^{2}+\frac{k}{a^{2}}\Big).
\end{equation}
Similarly, for space components we have
\begin{equation}
{G^{i}}_{j}=\frac{{\Pi^{i}}_{j}}{m_{4}^{6}}-{{\cal{E}}^{i}}_{j}
\end{equation}
where
\begin{equation}
{G^{i}}_{j}=-\Big(2\dot{H}+3H^{2}+\frac{k}{a^{2}}\Big){\delta^{i}}_{j}.
\end{equation}
Now, using equation (22) we find \,
${\Pi^{0}}_{0}=-\frac{1}{12}\Big({{\Upsilon}^{0}}_{0}\Big)^{2}$ \,
and\,
${\Pi^{i}}_{j}=-\frac{1}{12}{{\Upsilon}^{0}}_{0}\Big({{\Upsilon}^{0}}_{0}-
2{{\Upsilon}^{1}}_{1}\Big){\delta^{i}}_{j}$. Also, the time and
space components of the total energy-momentum tensor are given by
\begin{equation}
{{\Upsilon}^{0}}_{0}=-\rho_{tot}-m_{3}^{2}{G^{0}}_{0}
\end{equation}
and
\begin{equation}
{{\Upsilon}^{i}}_{j}=-p_{tot}\delta^{i}_{j}-m_{3}^{2}{G^{i}}_{j},
\end{equation}
where $\rho_{tot}$ and $p_{tot}$ are given by (14) and (15). These
equations lead us to the following effective Friedmann equation on
the brane
\begin{equation}
3\bigg(H^{2}+\frac{k}{a^{2}}\bigg)={{\cal{E}}^{0}}_{0}+
\frac{1}{12m_{4}^{6}}\bigg[\rho_m-3\beta
H^2+\frac{1}{2}\dot{\phi}^{2}+n^{2}V(\phi)-3m_{3}^{2}\Big(H^{2}+\frac{k}{a^{2}}\Big)\bigg]^{2}.
\end{equation}
This equation contains Lorentz invariance violation via existence of
$\beta$. One can easily show the possibility of existence of two
different branches of this DGP-inspired Friedmann equation. On the
other hand, as the tensor structure in the action is very different
from the original DGP model, we can expect the above action contains
no ghost. In fact, we have many parameters here, hence there is a
chance to remove the ghost instabilities from our model. As we will
show, by suitable fine-tuning and constraining parameters of this
model with observational data, this model has the capability to
explain late-time acceleration and other cosmologically interesting
issues. For future reference, we note that using Codazzi equation
one can show that $\nabla^{\nu}{\cal{E}}_{\mu\nu}=0$ \,[21].\\

\section{Cosmological Aspects of the Model}
Late-time positively accelerated expansion and transition to phantom
phase of scalar field as a dark energy component are two interesting
challenges of modern cosmology. Therefore, in which follows, we
study late-time acceleration and dynamics of equation of state
parameter, $\omega_{\phi}(t)$ in this Lorentz violating DGP setup.
The equation of state parameter of scalar field on the brane is
given by
\begin{equation}
\omega_\phi=\frac{p_\phi}{\rho_\phi}=\Big[\frac{\frac{1}{2n^{2}}\dot{\phi}^{2}-V(\phi)}
{\frac{1}{2}\dot{\phi}^{2}+n^{2}V(\phi)}\Big]_{y=0}
\end{equation}
where from now on we set $n(0,t) = 1$. Dynamics of
$\omega_{\phi}(t)$ can be obtained in two different viewpoints.
Firstly, Friedmann equation (30) in the absence of ordinary matter
on the brane and with flat spatial geometry ($k=0$), can be
rewritten as follows
\begin{equation}
H^{2}=\frac{{{\cal{E}}}_{0}}{3a^4}+
\frac{1}{36m_{4}^{6}}\bigg[-3\beta
H^2+\frac{1}{2}\dot{\phi}^{2}+V(\phi)-3m_{3}^{2}H^{2}\bigg]^{2}.
\end{equation}
We take $\dot{{{\cal{E}}}^{0}}_{0}+4H{{\cal{E}}^{0}}_{0}=0$
\,\,which an integration gives
${{\cal{E}}^{0}}_{0}=\frac{{{\cal{E}}_{0}}}{a^{4}}$ \, where
${\cal{E}}_{0}$ is integration constant [21]. Dynamics of scalar
field can be deduced from this equation directly
\begin{equation}
\dot{\phi}^2=6H^2(\beta+m_3^2)-2V(\phi)+12m_4^3\epsilon\sqrt{H^2+\frac{{{\cal{E}}}_{0}}{3a^4}}
\end{equation}
where $\epsilon=\pm1$, corresponding to two possible embedding of
DGP braneworld. To determine exact dynamics of scalar field, we need
to specify functional form of potential, $V(\phi)$. In this
framework, we use two well-known potentials: \,
$V(\phi)=\lambda'\phi^2$\, and an exponential potential as \,
$V(\phi)=V_{0}\exp\bigg(-\sqrt{\frac{16\pi}{pm^2_{pl}}}\phi\bigg)$
[22].

Secondly, we see that equation (33) depends on the potential of
scalar field. With this facility, we can obtain a
potential-independent equation of dynamics for scalar field. The
energy equation for vector field $u$ is
\begin{equation}
\dot{\rho}_u+3H(\rho_u+p_u)=+3H^2\dot{\beta}
\end{equation}
and for the scalar field we find
\begin{equation}
\dot{\rho}_\phi+3H(\rho_\phi+p_\phi)=-3H^2\dot{\beta}.
\end{equation}
There is a non-conservation scheme in this setup due to
energy-momentum transfer between scalar and vector fields. This is
very similar to the case studied by Zimdahl {\it et al} [23]. As
they have shown, a coupling between a quintessence scalar field and
a cold dark matter (CDM) fluid leads to a stable, constant ratio for
the energy densities of both component compatible with a power law
accelerated cosmic expansion. In fact this coupling is responsible
for accelerated expansion and possible crossing of phantom divide
line. In our Lorentz invariance violating scenario this coupling is
present between scalar field and vector field leading to an {\it
interactive} picture. Note that one can consider also an interaction
between ordinary matter and vector field on the brane. Very
recently, authors of Ref. [24] have investigated the cosmological
evolution of an interacting scalar field model in which the scalar
field has an interaction with the background matter via
Lorentz violation in $4$-dimensional model.\\
Nevertheless, the total energy in the presence of both scalar and
vector fields is conserved
\begin{equation}
\dot{\rho}+3H(\rho+p)=0,~~~ (\rho=\rho_u+\rho_\phi).
\end{equation}
This energy conservation equation can also be obtained by equating
the covariant divergence of the total energy-momentum tensor to
zero, since the covariant divergence of the Einstein tensor is zero
by its geometric construction. It follows from contraction of the
geometric Bianchi identity.\\
We obtain dynamics of the scalar field by differentiating equation
(16) with respect to $t$ and then using equation (35) to find
\begin{equation}
\ddot{\phi}+3H\dot{\phi}+3H^2\beta_{,\phi}+V_{,\phi}=0.
\end{equation}
By differentiating equation (32) with respect to $t$ and using
equation (37) we have
\begin{equation}
\dot{\phi}=\mp2m_4^3(\frac{2H_{,\phi}}{\dot{\phi}H}+\frac{12{\cal{E}}_{0}}{\dot{\phi}^2\dot{a}a^3})^{\frac{1}{2}}
-2H\beta_{,\phi}-2\beta H_{,\phi}-m_3^2\frac{H_{,\phi}}{H}
\end{equation}
where we assume $ H $ and $\beta $ are depended on $\phi$ in
forthcoming arguments and for simplicity, we set ${\cal{E}}_{0}=0$.
By substituting equation (38) into the Friedmann equation (32), the
potential of the scalar field in our model takes one of the
following forms
\begin{equation}
V(\phi)=\frac{1}{2}\frac{12H+6{
m_3}^{2}{H}^{2}{m_4}^{3}-{\dot{\phi}}^{2}{{m_4}^{3}+6{H}^{2}{m_4}^{3}\beta}}{{
m_4}^{3}}
\end{equation}
and
\begin{equation}
V(\phi)=\frac{1}{2}\frac{-12H+6{
m_3}^{2}{H}^{2}{m_4}^{3}-{\dot{\phi}}^{2}{{m_4}^{3}+6{H}^{2}{m_4}^{3}\beta}}{{
m_4}^{3}}
\end{equation}
where $\dot{\phi}$ can be obtained from equation (38). We emphasize
that equation (38) has several solutions, here we just consider one
real root of this equation to proceed further. On the other hand,
equation governing on dynamics of $\phi$ itself has no analytical
solution. So, one can try to find some intuition by numerical
analysis of the parameter space of the model.

To obtain dynamics of equation of state parameter, we note that
there are two governing equations on the dynamics of $H^2$ as
follows
\begin{equation}
H^2=\frac{1}{3}\,{\frac
{\rho_{\phi}\,{m_3}^{2}+\beta\,\rho_{\phi}+6\,{m_4}^{6}+2\,\sqrt
{3\,{m_4}^{6}
\rho_{\phi}\,{m_3}^{2}+3\,{m_4}^{6}\beta\,\rho_{\phi}+9\,{m_4}^{12}}}{2\,\beta\,{m_3}^{2}+{
\beta}^{2}+{m_3}^{4}}}
\end{equation}
and
\begin{equation}
H^2=\frac{1}{3}\,{\frac
{\rho_{\phi}\,{m_3}^{2}+\beta\,\rho_{\phi}+6\,{m_4}^{6}-2\,\sqrt
{3\,{m_4}^{6}
\rho_{\phi}\,{m_3}^{2}+3\,{m_4}^{6}\beta\,\rho_{\phi}+9\,{m_4}^{12}}}{2\,\beta\,{m_3}^{2}+{
\beta}^{2}+{m_3}^{4}}}
\end{equation}
Using these two branches of the model and also equation (35), we
find
\begin{equation}
E_{1}+E_{2}(1+\omega_{\phi})=-\dot{\beta}E_{3}
\end{equation}
where we have defined
\begin{equation}
E_{1}=\frac{\dot{\rho_{\phi}}}{\rho_{\phi}}={\frac
{{\dot{H}}}{H}}+{\frac {{\dot{\beta}}\,H{{m_4}}^{3}+\beta
\,{\dot{H}}\,{{m_4}}^{3}+{{m_3}}^{2}{\dot{H}}\,{{m_4}
}^{3}}{2\,{\epsilon_4}+\beta\,H{{m_4}}^{3}+{{m_3}}^{2}H{{m_4}}^{3}}},
\end{equation}

\begin{equation}
E_{2}=\frac{3\,{\epsilon_1}\, \left( {{m_4}}^{3}+{\epsilon_2}\,\sqrt
{{\frac {{{ m_4}}^{9}+{H}^{2} \left( \beta+{{m_3}}^{2} \right)
^{2}{{ m_4}}^{3}+2\,H{\epsilon_4}\, \left( \beta+{{m_3}}^{2} \right)
}{{{m_4}}^{3}}}} \right)} { \left( \beta+{{m_3}}^{2} \right)},
\end{equation}
and
$$E_{3}=\frac {2\,{\epsilon_3}\,\sqrt {{{m_4}}^{3} \left(
{{m_4}}^{9} +{H}^{2} \left( \beta+{{m_3}}^{2} \right)
^{2}{{m_4}}^{3}+2 \,H{\epsilon_4}\, \left( \beta+{{m_3}}^{2} \right)
\right) }{{m_4}}^{3}}{H \left( {{m_4}}^{3} \left( \beta+{{m_3}}^ {2}
\right) H+2\,{\epsilon_4} \right) \left( \beta+{{m_3}}^{2}
 \right) ^{2}}$$
\begin{equation}
+\frac{{H}^{2} \left( \beta+{{m_3}}^{2} \right)
^{2}{{m_4}}^{3}+2\,H{\epsilon_4}\, \left( \beta+{{m_3}}^{2}\right)
+2\, {{m_4}}^{9}}{H \left( {{m_4}}^{3} \left( \beta+{{m_3}}^ {2}
\right) H+2\,{\epsilon_4} \right) \left( \beta+{{m_3}}^{2}
 \right) ^{2}}.
\end{equation}
In these relations we have defined
$\epsilon_{1}=\epsilon_{2}=\epsilon_{3}=\epsilon_{4}=\pm 1$ where
have been appeared due to algebraic structure of the model. Although
all of these quantities have the same value, we have to save them in
forthcoming equations because of several permutations of signs in
equations of cosmological dynamics. In fact, as we will show,
suitable and simultaneous choices of these quantities are important
in analysis of parameters space. Now, equation of state parameter
takes the following complicate form
$$\omega_{\phi}=\frac{1}{6}\, \Bigg[ -6\,{\epsilon_2}\, \left(
\beta+{{m_3}}^{2} \right) H{ \epsilon_1}\, \left( 1/2\,{{m_4}}^{3}
\left( \beta+{{m_3}}^{2}
 \right) H+{\epsilon_4} \right)\times$$
 $$\sqrt {{\frac {{{m_4}}^{9}+{H}^{2}
 \left( \beta+{{m_3}}^{2} \right) ^{2}{{m_4}}^{3}+2\,H{\epsilon_4}\, \left( \beta+{{m_3}}^{2} \right)
 }{{{m_4}}^{3}}}}-$$
 $$2 \, {\beta_{,\phi}}\,{\dot{\phi}}\,{\epsilon_3}\,\sqrt
{{{m_4}}^{3}
 \left( {{m_4}}^{9}+{H}^{2} \left( \beta+{{m_3}}^{2}
 \right) ^{2}{{m_4}}^{3}+2\,H{\epsilon_4}\, \left( \beta+{{m_3
}}^{2} \right)  \right) }{{m_4}}^{3}-$$
$$3\,{{m_4}}^{3} \left( \beta+{{m_3}}^{2} \right) ^{2} \left(
2/3\,{\beta_{,\phi}}\,{\dot{\phi}}+{\epsilon_1}\,{{m_4}}^{3} \right)
{H}^{2}-$$
$$6\, \bigg(
 \left( 1/3\,{H_{,\phi}}\,{\dot{\phi}}\,{{m_3}}^{4}+2/3\,{H_{,\phi}
}\,{\dot{\phi}}\,\beta\,{{m_3}}^{2}+{\epsilon_1}\,{\epsilon_4}+1/3\,{
\beta}^{2}{H_{\phi}}\,{\dot{\phi}} \right)
{{m_4}}^{3}+1/3\,{\beta_{,\phi}}\,{\dot{\phi}}\,{\epsilon_4} \bigg)
\left( \beta+{{m_3}}^ {2} \right) H-$$
$$2\, \left( {{m_4}}^{9}{\beta_{,\phi}}+{H_{,\phi}}\,{ \epsilon_4}\,
\left( \beta+{{m_3}}^{2} \right) ^{2} \right) {\dot{\phi}} \Bigg]
\Bigg[\left( {{m_4}}^{3}+{\epsilon_2}\,\sqrt {{\frac {{{
m_4}}^{9}+{H}^{2} \left( \beta+{{m_3}}^{2} \right) ^{2}{{
m_4}}^{3}+2\,H{\epsilon_4}\, \left( \beta+{{m_3}}^{2} \right)
}{{{m_4}}^{3}}}} \right)^{-1}$$
\begin{equation}
\times\left( \beta+{{m_3}}^{2}
 \right) ^{-1}{H}^{-1}{{\epsilon_1}}^{-1} \left( 1/2\,{{m_4}}^{3}
 \left( \beta+{{m_3}}^{2} \right) H+{\epsilon_4} \right)
 ^{-1}\Bigg].
\end{equation}
This equation will be used to perform numerical analysis of the
model. As another important cosmological parameter, the deceleration
parameter $q$ which is defined as
\begin{equation}
q=-\frac{\ddot{a}a}{\dot{a}^2}=-1-\frac{\dot{H}}{H^2}
\end{equation}
using equation (32), takes the following form
$$q=\frac{1}{3}\Bigg[
{-2\sqrt{6}\epsilon_5}\sqrt{\Big(6{{m_4}}^{6}{H}^{4}+
\left(2H\beta+{{m_3}}^{2}\right)\left(-3{\dot{\beta}}{H}^{2}+{\dot{V}(\phi)}+{\ddot{\phi}}{\dot{\phi}}
 \right)\Big){H}^{4}{{m_4}}^{6}}+$$
$$ \left(-12{\beta}^{2}-12 {{m_4}}^{6}\right){H}^{4}-12\beta\left(
{{m_3}}^{2}
-1/2{\dot{\beta}}\right){H}^{3}+\left(-3{{m_3}}^{4}+3\,{\dot{\beta}}{{m_3}}^{2}
\right){H}^{2}-2\beta\left({\dot{V}(\phi)}+{\ddot{\phi}}{
\dot{\phi}}\right)H$$
\begin{equation}
-{{m_3}}^{2}
\left({\dot{V}(\phi)}+{\ddot{\phi}}{\dot{\phi}}\right)\Bigg]\Big[{\left(
2H\beta+{{m_3}}^{2}\right)^{2}{H}^{2}}\Big]^{-1}.
\end{equation}
Recent observations of distant type Ia supernovae and other
observational data [9] indicate that $q$ is currently negative; that
means the expansion of the universe is positively accelerated. This
is an indication that the gravitational attraction of matter, on the
cosmological scale, is more than contraction by negative pressure
dark energy in the form of quintessence. In which follows, we study
cosmological implications of our model focusing on late-time
acceleration and scalar field dynamics as a candidate of dark energy
and possible transition to phantom phase. We show that the Lorentz
symmetry breaking scalar field can be treated as a good candidate
for the role of the dark energy source.\\
Before proceeding further, we should address some important and
related issues here. Firstly, one should be careful to choose the
appropriate equation of state for components that are used to
describe the universe energy-momentum content. As we have emphasized
earlier, a suitable coupling between a quintessence scalar field and
other matter content can leads to a constant ratio of the energy
densities of both components which are compatible with an
accelerated expansion of the universe or crossing of the phantom
divide line (for more details see [23] and reference therein). In
this respect and for instance, the holographic dark energy models
studied in Ref. [7] have the phantom phase by adopting a native
equation of state, whereas the authors in [25] have found
accelerating phase only using the effective equation of state.\,
Based on these arguments, we should explain what kind of equation of
state is used for observing the nature of mixed fluids here. In our
model, we have three sources of energy-momentum: 1- standard
ordinary matter, 2- scalar field as a candidate of dark energy and
3- energy-momentum content depended on Lorentz violating vector
field. Here we assume that standard matter has negligible
contribution on the total energy-momentum content of the universe
and we can consider a constant linear isothermal equation of state
as $p_m=(\gamma_{m}-1)\rho_m$ that $1\leq\gamma_{m}\leq 2$ for it.
For other two energy-momentum contents, it is possible to use the
"trigger mechanism" to explain dynamical equation of state [26].
This means that we assume scalar- vector-tensor theory containing
Lorentz invariance violation which acts like the hybrid inflation
models. In this situation, vector and scaler field play the roles of
inflaton and the "waterfall" field respectively [26]. In this
regard, we can fine-tune parameter $m$ and other parameters to
obtain best fit model using the observational data. Of course, an
attractor solutions and fine-tuning in Lorentz violation model for
suitable inflation phase has been studied in Ref. [5]. Therefore it
is reasonable to expect that one of them will eventually dominate to
explain inflation or accelerating phase and crossing of phantom
divide line. We should emphasize that the model studied in this
paper belongs to a wider class of Lorentz-violating theories
exhibiting the phantom behavior (see for instance Refs. [4,5,27] for
a number of Lorentz-violating models). One can extend this framework
to understand issues such as transient phantom stage, super-horizon
ghosts and to deal with the question that how generic are the
features obtained in this particular model of late-time de Sitter
attractor. Some of these issues have been discussed in Ref. [28].
One more direction in this framework is to modify our model in such
a way that it find the capability to describe inflationary epoch
rather than the late-time acceleration. As has been pointed in [28],
this type of model may give rise to some distinct features in the
CMB spectrum.

\section{Numerical Analysis of the Parameters Space}
In this section we study late-time acceleration and possible
transition to phantom phase of the scalar field. Incorporation of
the Lorentz invariance violation in the model provides a wider
parameter space (relative to the case with one Lorentz preserving
scalar field) which leads to more suitable framework for explanation
of these interesting cosmological aspects. To show the validity of
this statement, we need to solve equations (47) and (49). In the
first stage, we obtain dynamics of scalar field $\phi$ using
equation (33). This goal will be achieved only if the Hubble
parameter $H(\phi(t))$ and the vector field coupling,\,
${\beta}(\phi(t))$ are known a priori. In which follows, our
strategy is to choose some suitable and natural candidates for the
Hubble parameter $H(\phi(t))$ and the vector field coupling
$\beta(\phi(t))$. Then we focus on possible crossing of the phantom
divide line and realization of the universe late-time acceleration.
We obtain suitable domains of parameters space which admit late-time
acceleration and crossing of the phantom divide line by equation of
state parameter. Probably this Lorentz violating DGP-inspired model
has some important consequences in the spirit of cosmology ( such as
possible realization of bouncing solutions) and particle physics,
but here we focus only on late-time acceleration and transition to
phantom phase of the scalar field.

We consider a general case where both the vector field coupling and
the Hubble parameter are functions of scalar field $\phi$ defined as
follows
\begin{equation}
H=H_0\phi^{\zeta} \ , \quad  \beta(\phi) = m\phi^\xi
\end{equation}
where $H_0$ and $m$ are positive and constant parameters. Note that
scalar field itself is depended on the cosmic time, $t$. In this
case, dynamics of scalar field with potential of the type
$V(\phi)=\lambda'\phi^2$ is given by
\begin{equation}
\dot{\phi}=6{H_0}^2\phi^{2\zeta}(m\phi^{\xi}+{m_3}^2)-\lambda'\phi^2+12{m_4}^3\epsilon{H_0}\phi^{\zeta}
\end{equation}
and for potential of the type
$V(\phi)=V_{0}\exp\bigg(-\sqrt{\frac{16\pi}{pm^2_{pl}}}\phi\bigg)$,
we find
\begin{equation}
\dot{\phi}=6{H_0}^2\phi^{2\zeta}(m\phi^{\xi}+{m_3}^2)-
V_{0}\exp\bigg(-\sqrt{\frac{16\pi}{pm^2_{pl}}}\phi\bigg)+12{m_4}^3\epsilon{H_0}\phi^{\zeta}.
\end{equation}
The question then arises here: what choices of space parameters lead
to analytical solutions of the equations (51) and (52)? The answer
to this question is summarized in tables $1$ and $2$. We note that
as table $2$ shows, for exponential potential there is very limited
possibility to obtain analytical solution for scalar field dynamics.
Nevertheless, we can find analytical solution in some especial
cases. For instance, form equation (51) with $\xi=-1$ and $\zeta=1$
we find the following analytical solution for dynamics of scalar
filed
$$\phi(t)
=\frac{1}{2}\Bigg[\Bigg(36{{m_4}}^{6}{\epsilon}^{2}{{H_0}}^{2}{e^{2{A_0}\sqrt
{6{{H_0}}^{2}{{m_3}}^{2}-\lambda'}}}+9{{H_0}}^{4}{m}^{2}{e^{2{A_0}\sqrt
{6{{H_0}}^{2}{{m_3}}^{2}-\lambda'}}}-$$
$$12{e^{\left(t+{A_0}\right)\sqrt{6{{H_0}}^{2}{{ m_3}}^{2}-\lambda'}
}}\sqrt{6{{H_0}}^{2}{{m_3}}^{2}-\lambda'}{{m_4}}^{3}\epsilon{H_0}+6{e^{2t\sqrt
{6\,{{H_0}}^{2}{{m_3}}^{2}-\lambda'}}}{{H_0}}^{2}{{m_3}}^{2}-$$
$${e^{2t\sqrt {6{{H_0}}^{2}{{m_3}}^{2}-\lambda'}}}\lambda'-
6{e^{\left(t+{A_0}\right)\sqrt{6{{H_0}}^{2}{{m_3}}^{2}-\lambda'}}}
\sqrt{6{{H_0}}^{2}{{m_3}}^{2}-\lambda'}{{H_0}}^{2}m+36{{H_0}}^{3}m{{m_4}}^{3}\epsilon{e^{2{A_0}\sqrt
{6{{H_0}}^{2}{{m_3}}^{2}-\lambda'}}}
 \Bigg)$$
\begin{equation}
{e^{-\left(t+{A_0}\right)\sqrt{6{{H_0}}^{2}{{m_3}}^{2}-\lambda'}
}}\Bigg]\Big[\left(6{{H_0}}^{2}{{m_3 }}^{2}-\lambda'\right)
^{3/2}\Big]^{-1}
\end{equation}
where $A_0$ is an integration constant. We emphasize that this
analytical solution of equation (51) is obtained under some especial
choices of parameter space. Without these choices it is impossible
to find closed analytical solution for $\phi(t)$. We use this
solution for our forthcoming numerical analysis. Figure $1$ shows
dynamics of $\phi$ as described by equation (53) for two different
cases.

\begin{figure}[htp]
\begin{center}\includegraphics{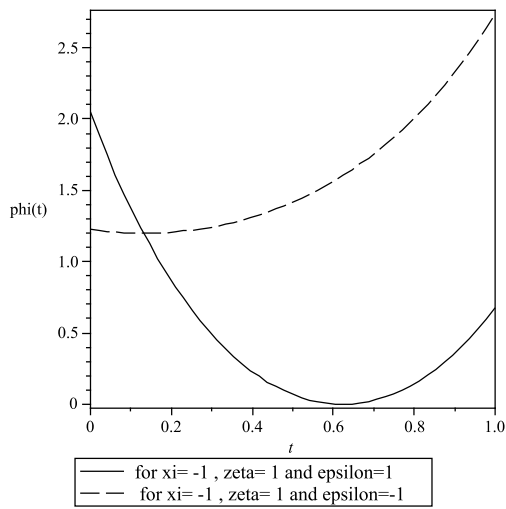} \vspace{5.9cm}
\end{center}
 \caption{\small {Dynamics of scalar field in two especial cases.}}
\end{figure}

\begin{table}
\begin{center}
\caption{Acceptable range of $\xi$  and $\zeta$ to have analytical
solution of the scalar field equation (51)} \vspace{0.5 cm}
\begin{tabular}{|c|c|c|c|c|c|c|c|}
  \hline
  \hline values of $\xi$ and $\zeta$ & analytical solution for equation (51)?     \\
  \hline  $\xi\neq 0$ and $\zeta<0$ & no \\
  \hline $\xi=0$ and $\zeta=0$ & yes \\
  \hline $\xi=1$ and $\zeta=0$ & yes \\
  \hline $\xi=2$  and $\zeta=0$ & yes \\
  \hline $\xi\geq1$  and $\zeta\geq1$ & no \\
  \hline $\xi=-1$ and $\zeta=1$ & yes \\
  \hline $\xi=-2$ and $\zeta=1$ & yes \\
  \hline other $\xi$ and $\zeta$ & should be examined \\
       \hline
\end{tabular}
\end{center}
\end{table}
\begin{table}
\begin{center}
\caption{Acceptable range of $\xi$  and $\zeta$ to have analytical
solution of the scalar field equation (52)} \vspace{0.5 cm}
\begin{tabular}{|c|c|c|c|c|c|c|c|}
  \hline
  \hline values of $\xi$ and $\zeta$ & analytical solution for equation (52)?     \\
  \hline  $\xi\neq 0$ and $\zeta\neq 0$ & no \\
  \hline $\xi=0$ and $\zeta=0$ & yes \\
  \hline $\xi=1$ and $\zeta=0$ & no \\
         \hline
\end{tabular}
\end{center}
\end{table}
Now by substituting equation (53) into equation (49), we can analyze
dynamics of deceleration parameter $q$ with respect to cosmic time
$t$. This has been shown in figures $2$ and $3$ and corresponding
results are summarized in table $3$.
\begin{figure}
\begin{center}\includegraphics{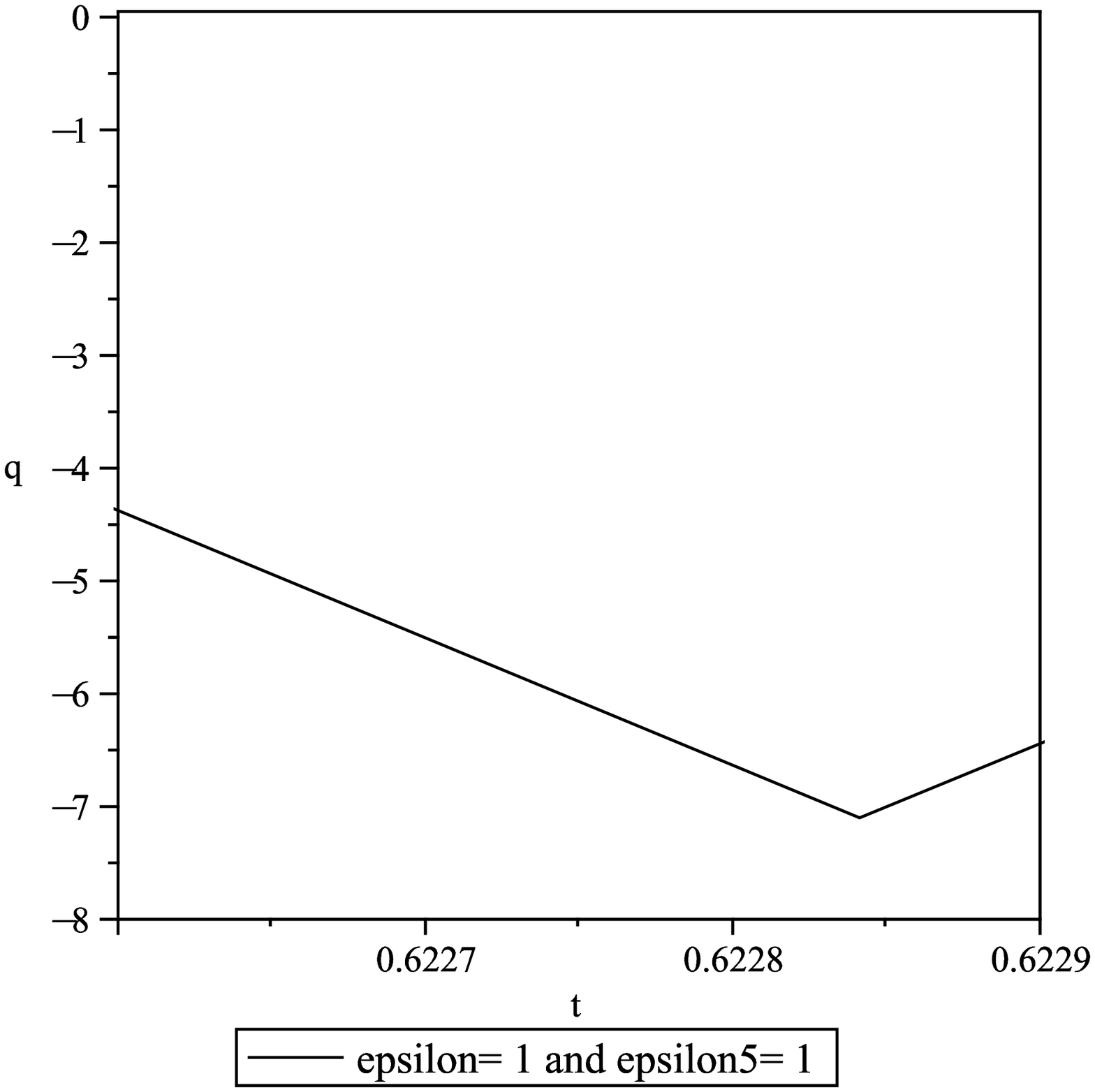} \vspace{5.5cm}\includegraphics{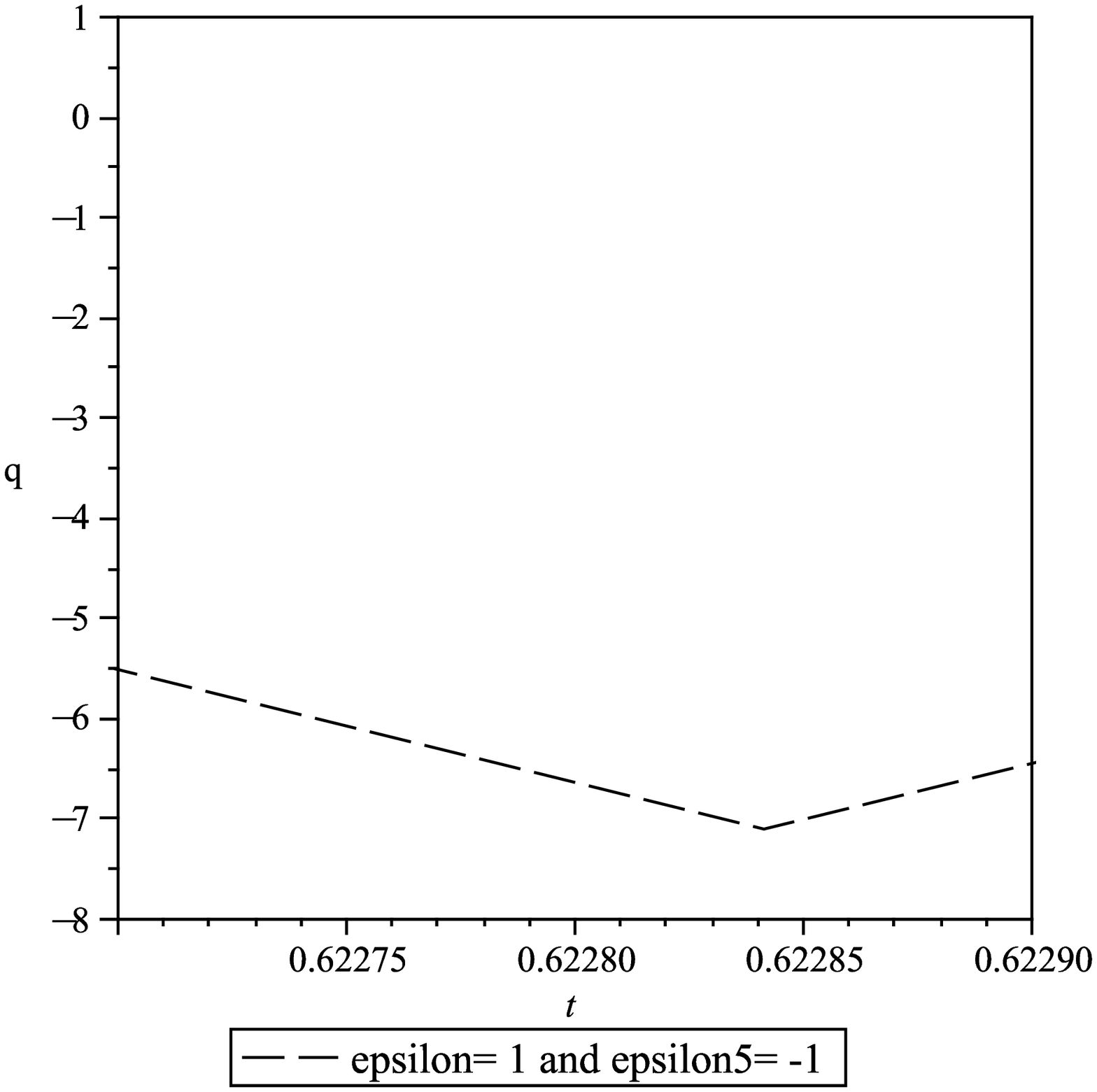}
\end{center}
\caption{\small {Variation of deceleration parameter $q$ relative to
cosmic time $t$ for $\epsilon=+1$ and $\epsilon_5=\pm 1$ with scalar
field potential of the type $V(\phi)=\lambda'\phi^2$.}}
\end{figure}

\begin{figure}[htp]
\begin{center}\includegraphics{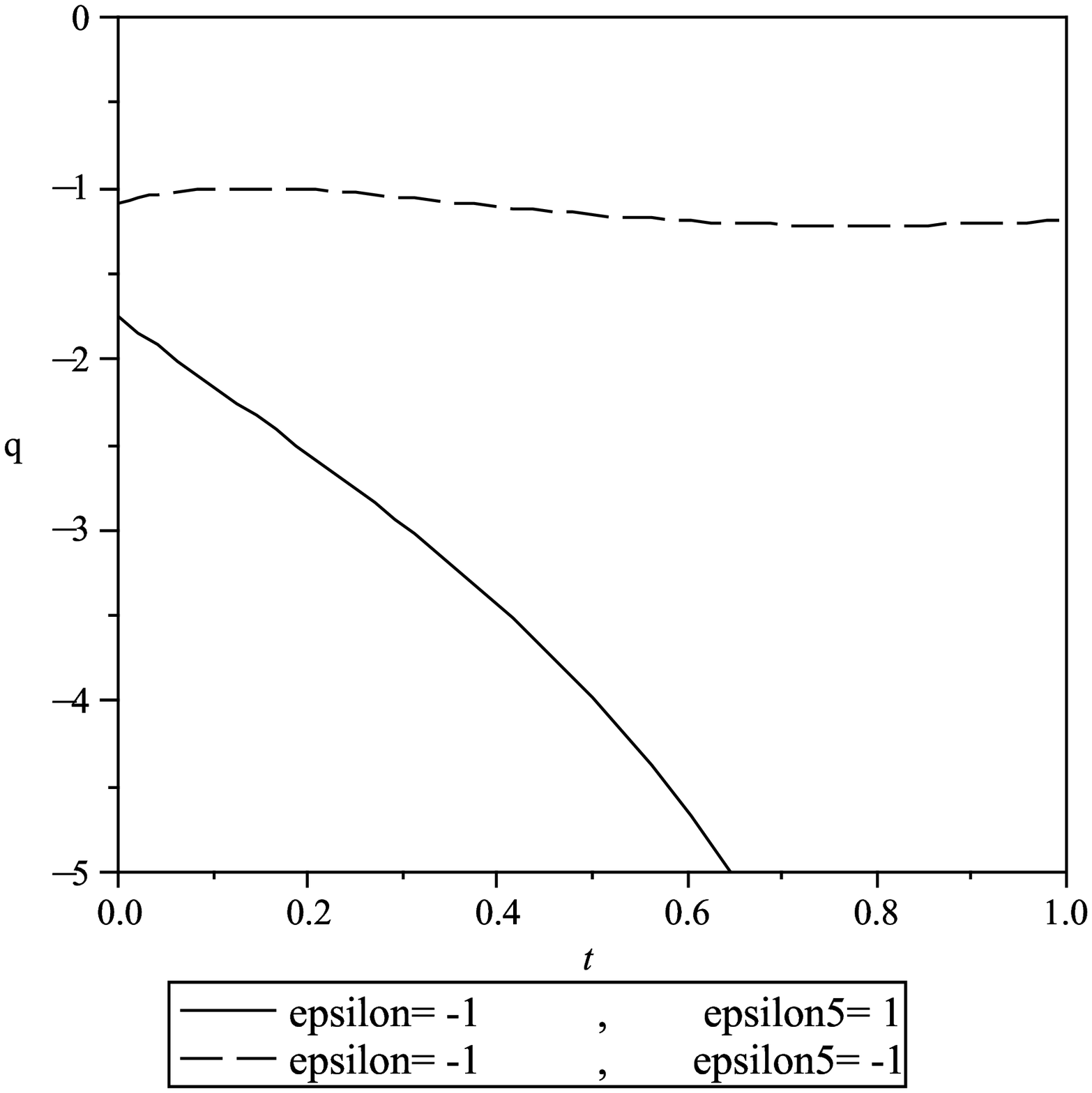} \vspace{5.9cm}
\end{center}
 \caption{\small {Variation of deceleration parameter $q$ relative to
cosmic time $t$ for $\epsilon=-1$ and $\epsilon_5=\pm 1$ with scalar
field potential of the type $V(\phi)=\lambda'\phi^2$.}}
\end{figure}

\begin{table}
\begin{center}
\caption{Summary of results from figures $2$ and $3$.} \vspace{0.5
cm}
\begin{tabular}{|c|c|c|c|c|c|c|c|}
  \hline
  \hline $\epsilon$&$\epsilon_5$ &deceleration parameter $q$ & late-time acceleration ? \\
  \hline +1&+1& negative & yes \\
  \hline +1&-1& negative & yes \\
  \hline -1&+1& negative & yes \\
  \hline -1&-1& negative but almost constant & yes \\
       \hline
\end{tabular}
\end{center}
\end{table}
\newpage
In which follows, we consider equation of state as given by equation
(47) with equation (53) for dynamics of scalar filed in order to
study crossing of the phantom divide line in this setup. For this
purpose, we set $\xi=-1$ and $\zeta=1$ to be more specific. The
results of numerical calculations are shown in figures $4$, $5$, $6$
and $7$. All of these figures show the possibility of transition to
the phantom phase of the scalar field. It is interesting to note
that as these figures show, transition from quintessence to phantom
phase and from phantom phase to quintessence is possible in this
braneworld scenario with appropriate choices of model parameters.
Since one minimally coupled scalar field in the presence of Lorentz
invariance symmetry cannot realize phantom divide line crossing,
observation of this crossing in our model is a result of interactive
nature of the model due to Lorentz invariance violation.  This is
manifested in our model via existence of the term such as $-3\beta
H^{2}$ in Friedmann equation (32) or $-3\dot{\beta}H^{2}$ in
equation (35).

We should stress here that by neglecting ordinary matter content (
$\rho_{m}=0=p_{m}$) in equation (20), we find
\begin{equation}
\omega_{tot}=\bigg[\frac{\beta
H^2\Big[3+2\frac{\dot{H}}{H^2}+2\frac{\dot{\beta}}{H\beta}\Big]
+\frac{1}{2n^{2}}\dot{\phi}^{2}-V(\phi)}{-3\beta
H^2+\frac{1}{2}\dot{\phi}^{2}+n^{2}V(\phi)}\bigg]_{y=0}.
\end{equation}
The same procedure as described above, leads us to the dynamics of
$\omega_{tot}$ as shown in figure $8$. We see that for
$\epsilon=+1$,\, $\omega_{tot}$ never crosses the phantom divide
line.

\begin{figure}[htp]
\includegraphics{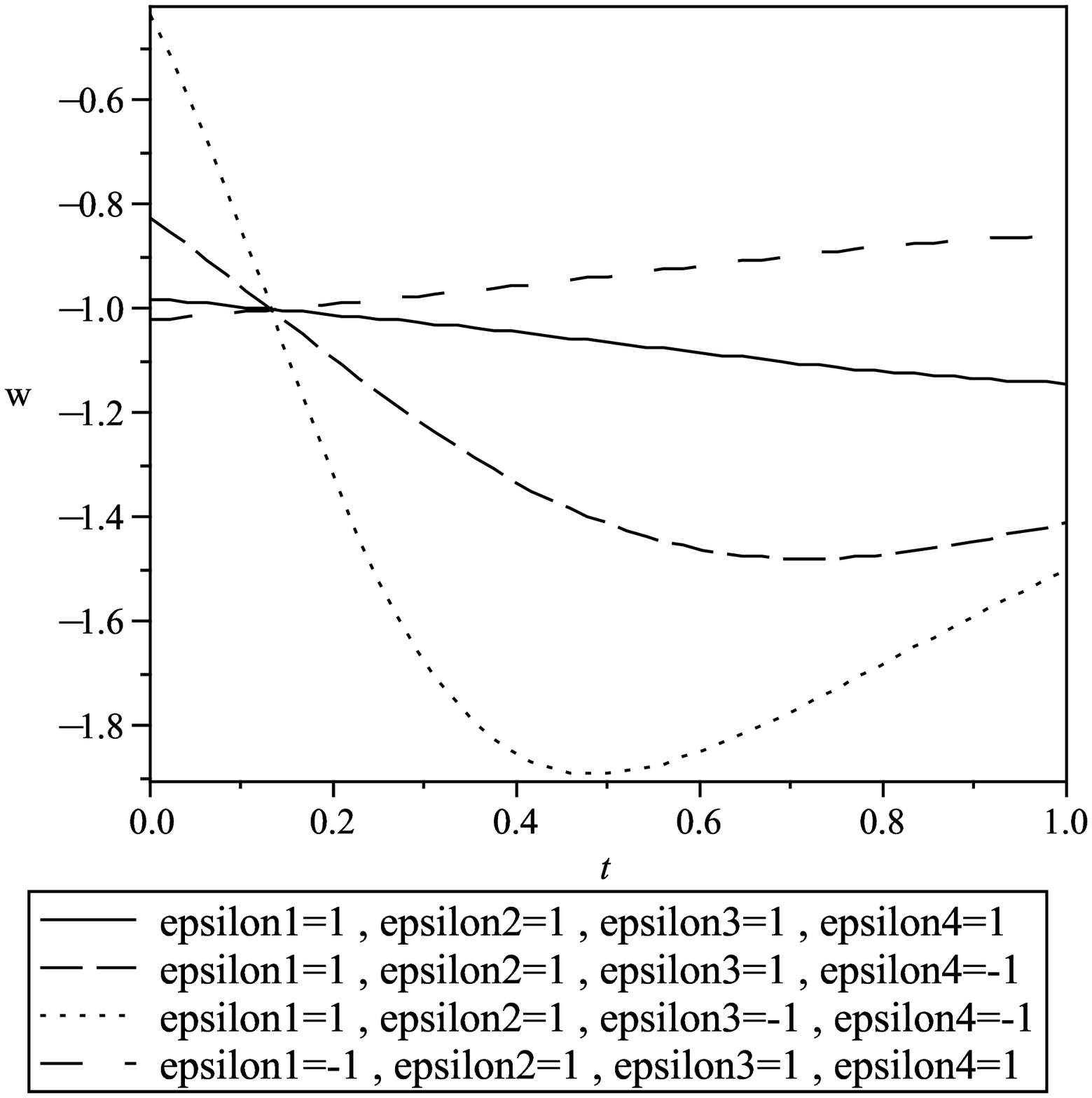} \vspace{7 cm}
 \caption{\small {Crossing of phantom divide line. }
 \hspace{11cm}}
 \vspace{10cm}

\includegraphics{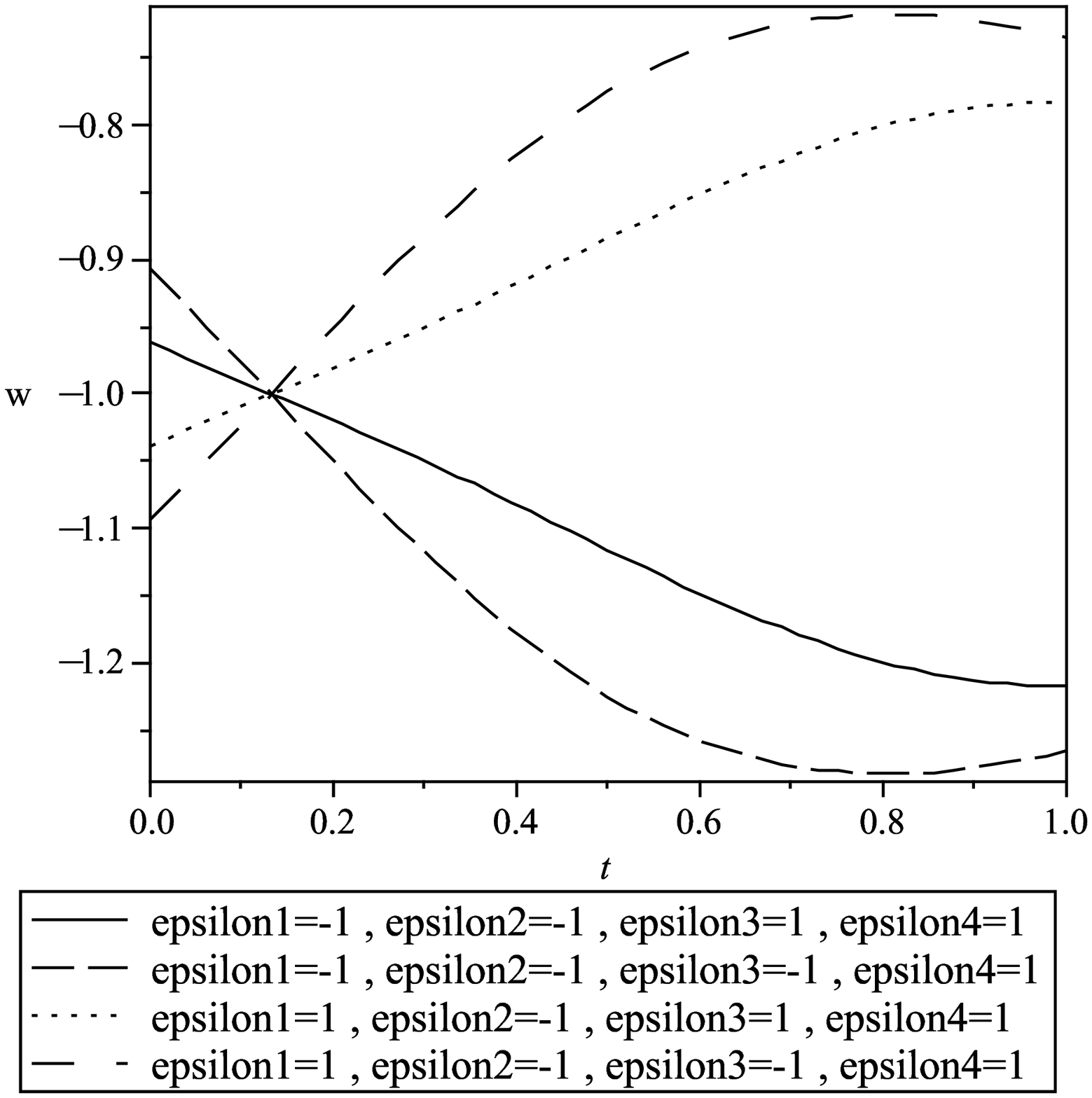} \vspace{-11 cm}  \caption{\small {Crossing the phantom
divide line.}\hspace{-12cm}}
  \vspace{10cm}

\includegraphics{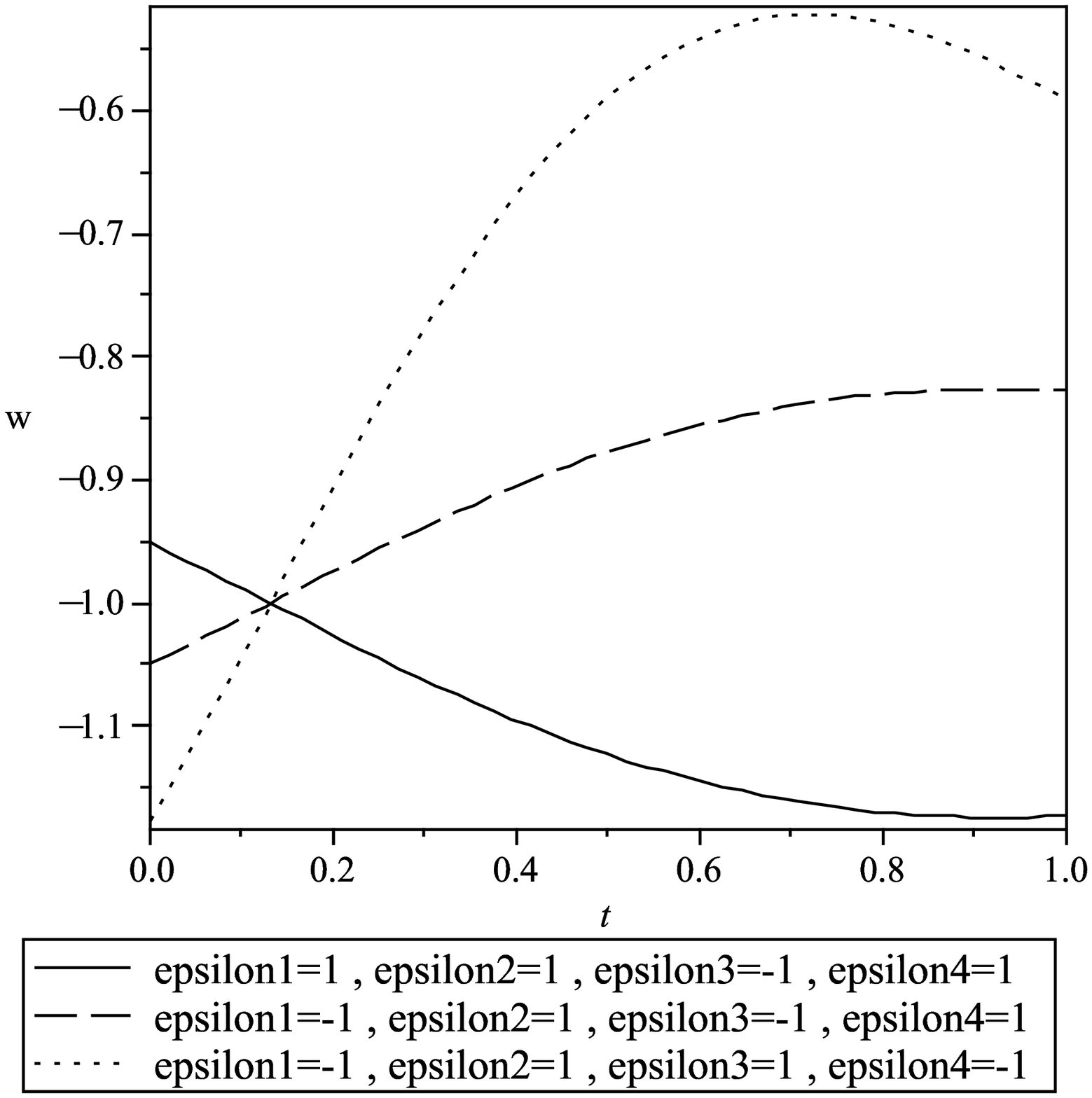}

\vspace{1 cm}
 \caption{\small {Crossing the phantom
divide line. }\hspace{11cm}}

\begin{center}
\includegraphics{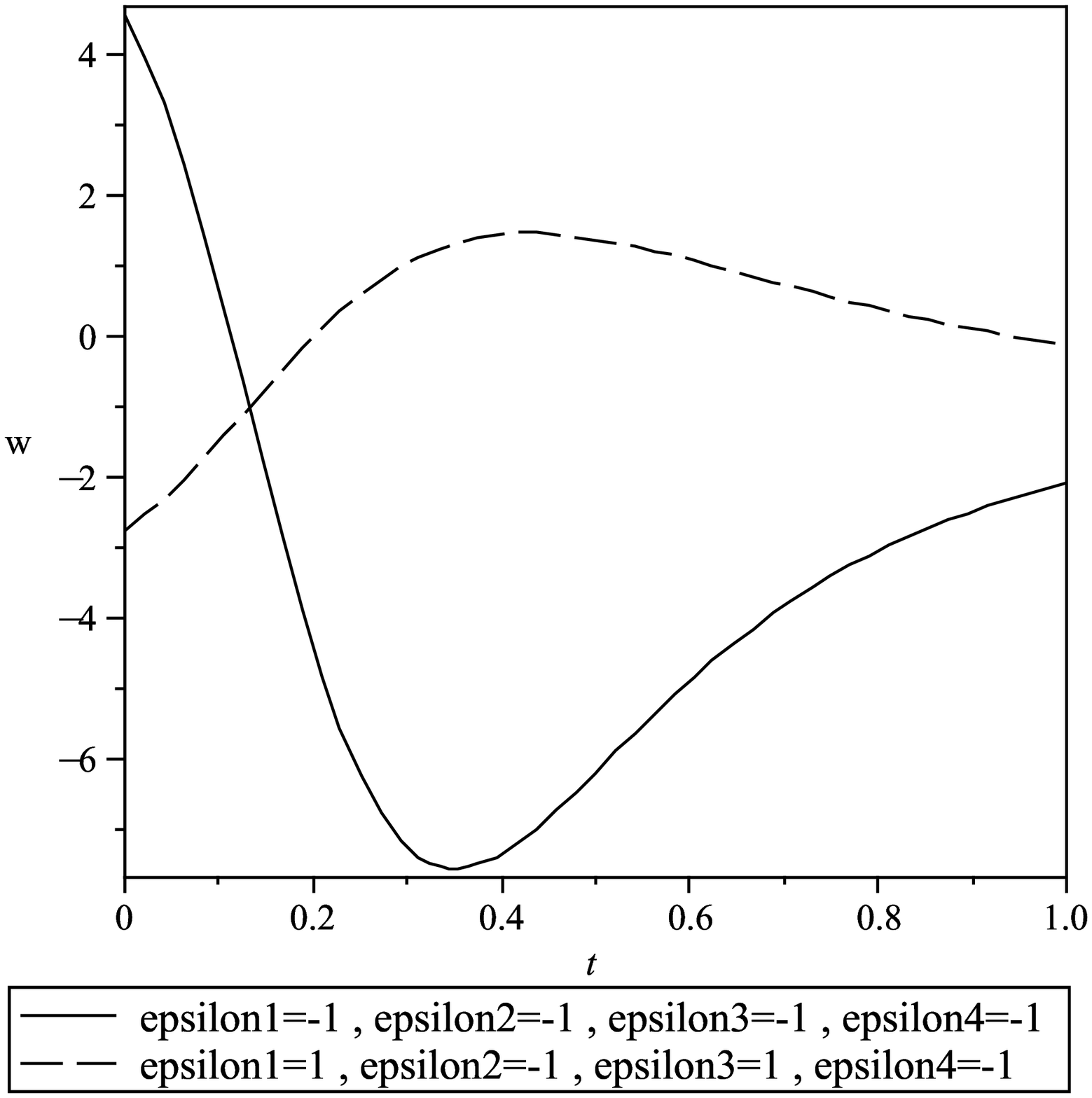}
\end{center}
\vspace{-2.5 cm}
 \caption{\small {Crossing the phantom
divide line.}\hspace{-12cm}}
 \end{figure}

\begin{figure}[htp]
\begin{center}\includegraphics{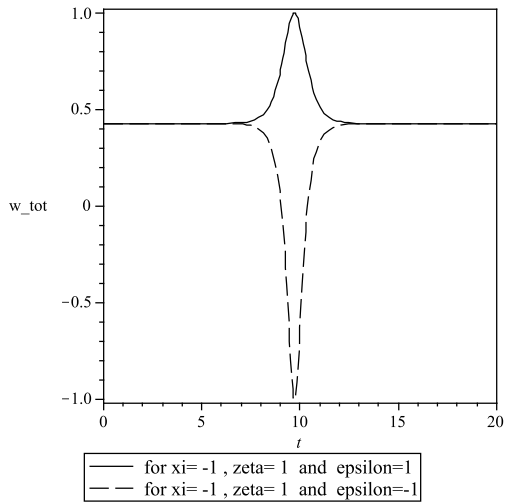} \vspace{5.9cm}
\end{center}
 \caption{\small {Dynamics of $\omega_{tot}$. There is no crossing for $\epsilon=+1$ branch of the model.}}
\end{figure}

\section{Summary and Discussion}
In this paper we have studied a DGP-inspired braneworld scenario
where the idea of Lorentz invariance violation has been incorporated
by specifying a preferred frame through the introduction of a
dynamical vector field normal to our brane. The Einstein field
equations obtained on the brane are modified by additional terms
emanating from the presence of the vector field. This model breaks
the $4D$ Lorentz invariance in the gravitational sector. As a new
mechanism for crossing of the phantom divide line by equation of
state parameter, we have shown that by a suitable choice of
parameters of the model, it is possible to have phantom divide line
crossing and realizing late-time acceleration in this Lorentz
invariance violating context.

More importantly, we need to show the stability of the solutions in
this Lorentz violating DGP-inspired model. It has been shown that
the self-accelerating branch of the DGP model contains a ghost at
the linearized level [29]. The ghost carries negative energy density
and it leads to the instability of the spacetime. The presence of
the ghost can be related to the infinite volume of the extra
dimension in the DGP setup. When there are ghost instabilities in
the self-accelerating branch, it is natural to ask what are the
results of solution decay. One possible answer to this question is
as follows: since the normal branch solutions are ghost-free, one
could think that the self-accelerating solutions may decay into the
normal branch solutions. In fact for a given brane tension, the
Hubble parameter in the self-accelerating universe is larger than
that of the normal branch solutions. Then it is possible to have
nucleation of bubbles of the normal branch in the environment of the
self-accelerating branch solution. This is similar to the false
vacuum decay in de Sitter spacetime. However, there are arguments
against this kind of reasoning which suggest that the
self-accelerating branch does not decay into the normal branch by
forming normal branch bubbles [29]. It was also shown that the
introduction of a Gauss–Bonnet term for the bulk does not help one
to overcome this problem [30]. In fact, it is still unclear what the
end state of the ghost instability is in the self-accelerated branch
of DGP inspired setups (for more details see [29]). On the other
hand, it seems that introduction of Lorentz violating vector field
on the brane provides a new degree of freedom which may provide a
suitable basis to treat ghost instability. Field theories of vector
fields are very restrictive and gauge invariant Lagrangians are the
ones that guarantee that the zeroth component of the gauge field is
not propagating and, therefore there are no ghosts. However, in the
action of our model as given by equation (3), the brane Lagrangian
for the vector field is not gauge invariant. Therefore, it is
possible that this vector field action by itself can introduce a
ghost, instead of curing the one of gravity. This is not necessarily
the case in the presence of vector field condensates [31,32]. On the
other hand, for certain timelike vector theories with spontaneous
Lorentz violation, suitable restrictions of the initial-value
solutions are identified that yield ghost-free models with a
positive Hamiltonian[33]. This restrictions can be imposed on
parameters such as $\beta_{i}$ in our model ( see also [19,30]).
Therefore, in our Lorentz invariance violating setup there is the
capability to overcome instabilities both in gravitational and
vector field sectors. Anyway, if the Lorentz violation annihilate
the ghost, this will be a great progress in the cosmology. We think
that our Lorentz invariance violating model on the DGP brane has the
capability to solve ghosts instabilities problem due to its wider
parameter space relative to other existing scenarios.

Within a similar viewpoint, very recently Zen {\it et al} [24] have
studied the cosmological evolution of an interacting scalar field
model in which the scalar field has its interaction with dark
matter, radiation, and baryon via Lorentz violation in $4D$ standard
model. They proposed a model of interaction through the effective
coupling parameter, $\bar{\beta}$,
$Q_m=-\frac{\dot{\bar{\beta}}\rho_m}{\bar{\beta}}$. They also
determined all critical points and studied their stability of
Lorentz violation model. On the other hand, Mariz {\it et al} [34]
studied the fact that the Lorentz-breaking parameter can be treated
as a natural explanation of the extremely small value of the
cosmological constant. Thus, the Lorentz symmetry breaking
introduces a mechanism for the arisal of a non-zero but very small
cosmological constant, and therefore providing an acceptable
solution for cosmological constant problem. We believe that Lorentz
symmetry breaking fields can be treated as a good candidate for the
role of the dark energy source and these types of models have the
capability to address other cosmological issues with better
adaptability than other models. Finally we should stress on some
open issues in this field: one issue concerns that we have provided
a scenario of a scalar field with Lorentz violation to realize the
crossing of the phantom divide. One maybe curious to know whether
this scenario can make some interesting predictions for
observations. For example the interactive terms in this model is
very important, but can this interaction affect the formation of
large scale structure, or the occupation of dark matter? This is a
very interesting issue, since as we know large scale structure and
the measure of dark matter are usually strongly constrained by CMB
observations. Moreover, we think this model is able to give a
bouncing solution of the universe. The reason lies in the fact that
as has been shown in Ref. [35], any models with crossing of phantom
divide line have the capability to realize a bouncing solution.\\

{\bf Acknowledgement}

It is a pleasure to thank Yi-Fu Cai for helpful discussions. We
would like also to thank an anonymous referee for his/her highly
valuable comments.

\end{document}